# Adsorption and dynamics of Si atoms at the monolayer Pb/Si(111) surface

Rakesh Kumar[a,b,c], Chuang-Kai Fang[c], Chih-Hao Lee[a], Ing-Shouh Hwang [c,*]

[a] *Department of Engineering and Systems Science, National Tsing Hua University, Hsinchu, Taiwan, Republic of China 30013*

[b] *Nano Science and Technology Program, Taiwan International Graduate Program, Academia Sinica, Nankang, Taipei, Republic of China 115*

[c] *Institute of Physics, Academia Sinica, Nankang, Taipei, Taiwan, Republic of China 115*

### ABSTRACT

In this work, we studied the adsorption behavior of deposited Si atoms along with their diffusion and other dynamic processes on a Pb monolayer-covered Si(111) surface from 125–230 K using a variable-temperature scanning tunneling microscope (STM). The Pb-covered Si(111) surface form a low-symmetry row-like ($\sqrt{7} \times \sqrt{3}$) structure in this temperature range and the Si atoms bind favorably to two specific on-top sites ($T_{1A}$ and $T_{1B}$) on the trimer row after deposition at the sample temperature of ~125 K. The Si atoms were immobile at low temperatures and started to switch between the two neighboring $T_{1A}$ and $T_{1B}$ sites within the same trimer when the temperature was raised to ~150 K. When the temperature was raised above ~160 K, the adsorbed Si atoms could hop to other trimers along the same trimer row. Below ~170 K, short hops to adjacent trimers dominated, but long hops dominated at temperatures above ~170 K. The activation energy and prefactor for the Si atoms diffusion were derived through analysis of continuous-time imaging at temperatures from 160–174 K. In addition, irreversible aggregation of single Si atoms into Si clusters started to occur at the phase boundaries or defective sites at temperatures above

~170 K. This study provides crucial information for understanding the very initial stage of nucleation and growth behavior of epitaxial Si layers on a Pb-covered Si(111) surface. In addition, our study provides strong evidence for breaking in the mirror symmetry in the ($\sqrt{7}\times\sqrt{3}$)-Pb structure, which has implication for the atomic model of this controversial structure.

# I. INTRODUCTION

The thin film growth in a layer-by-layer fashion is of great important to the modern semiconductor industry. For this purpose, surfactant mediated epitaxy (SME) [1-24] has been exploited in the hetero- and homo- epitaxial growth of both semiconductor and metal systems. In SME, a surfactant layer is deposited on the substrate of material "A" prior to growth of material "B". With the appropriate choice of this third material, the surfactant layer floats on top of the growing film during the deposition of material "B". Many SME systems have been reported to suppress 3D island growth, leading to layer-by-layer growth under appropriate conditions [16-19]. In spite of numerous studies, the way in which the surfactant improves the growth process remains unclear. Some studies suggest that the deposited atoms might be exchanged with the surfactant atoms soon after deposition, and thus their surface mobility is significantly reduced, resulting in a high island density compared to the case without the surfactant layer [3,4,5,6,15]. There are also other studies that demonstrate that the surface mobility of deposited atoms is significantly enhanced [17,21-23]. It is suggested that the surfactant atoms passivate the substrate surface dangling bonds; thus, the deposited atoms can move on the surfactant layer with small energy barriers [21-23].

Previous experimental studies of epitaxial growth usually focused on the formation of 2D islands and the growth of thin films; experiments were usually carried out at high deposition coverages and high temperatures when film growth can proceed. The atomic mechanisms of thin film growth were proposed based on indirect evidence and theoretical modeling. To elucidate the atomic mechanism in the nucleation and growth processes, it is necessary to investigate experimentally the processes directly at the very initial stage of deposition, such as which atomic sites individual atoms are adsorbed soon after deposition, and how deposited atoms diffuse, aggregate, etc. on the surface? This requires an atomic-scale study of the surfaces at very low deposition coverages and low temperatures. Such studies remain lacking. This is the main aim of the current study.

It has been demonstrated that one monolayer of Pb can serve as a good surfactant for the growth of Si and Ge thin films on Si(111) substrates [17,18,22-24]. The Pb coverage should be exactly one monolayer (1 ML = 1 monolayer = $7.84 \times 10^{14}$ atoms/cm$^2$), as determined by Rutherford backscattering (RBS) [14,18,24-28]; any excess or deficiency of Pb atoms cause defects in the grown thin films [18]. An early scanning tunneling microscope (STM) study indicated that the nucleation of two-dimensional (2D) Si epitaxial islands occurs only when the substrate temperature is sufficiently high and the Si deposition coverage is above a certain coverage [22]. At relatively low deposition coverages (down to ~0.03 ML) or low substrate temperatures (down to room temperature), the deposited Si atoms tend to self-assemble into a certain type of Si atomic wires [22]. The self-assembly of Si atomic wires at room temperature indicates that single Si atoms are sufficiently mobile at the Pb-covered Si(111) surface. To study the behavior of individual Si atoms at the very initial stage of deposition, we went down to ~120 K when the deposited Si atoms are not mobile on the surface. We then gradually increased the sample temperature and studied the adsorption and dynamics of individual Si atoms with an STM. Our results provide new insight into the mechanism in SME.

The monolayer Pb-covered Si(111) surface forms Si(111)-(1×1)-Pb [hereafter referred to as (1×1)-Pb] at room temperature (RT). When the sample is cooled down, the (1×1)-Pb structure is transformed reversibly into a low-symmetry row-like ($\sqrt{7} \times \sqrt{3}$)-Pb structure [34-38]. At ~125 K the adsorbed Si atoms are immobile and their adsorption sites are examined. Si atoms tend to appear near the on-top site ($T_1$ site) of the Si(111) substrate, but the adsorption of Si atoms exhibited a strong anisotropy among different $T_1$ sites, which has implications for the detailed atomic model of the ($\sqrt{7} \times \sqrt{3}$)-Pb structure. We will discuss two previously proposed models [34,38,41,42] that do not possess the breaking in the mirror symmetry uncovered here. The ($\sqrt{7} \times \sqrt{3}$)-Pb structure has received much attention because many interesting phenomena have been reported for this structure, such as a structural phase transformation [36,34,41,43], devil's staircase phase [29], superconducting transition [30,31,32,33], etc. To understand these phenomena, it is essential to determine the atomic structure of this phase.

## II. EXPERIMENTAL

Experiments were carried out using a commercial variable-temperature scanning tunneling microscope (Omicron VT-STM) in an ultrahigh vacuum chamber with a base pressure of ~6 x $10^{-11}$ mbar. First, a clean Si(111)-(7x7) surface was prepared [22,38-40].

A submonolayer of Pb (99.999% pure Pb) was evaporated from an effusion cell onto the Si(111)-(7x7) surface at RT, followed by a brief sample annealing at ~400 °C for 3-5 s. After annealing, individual domains of Si(111)-(1x1)-Pb were formed; each domain was surrounded by the Si(111)-(7x7) reconstruction. Some areas of the (1x1)-Pb surface were transformed gradually into a low-coverage Si(111)-($\sqrt{3} \times \sqrt{3}$)-Pb (Pb coverage ~ 1/3 ML) phase [22,40] if the annealing time was increased. The Pb coverage for the (1×1)-Pb phase on the bulk-terminated Si(111) substrate was determined by RBS to be one monolayer [14,18,24-28]. The prepared sample was transferred to the STM stage and cooled down to ~125 K. The temperature of the sample was measured with a Si diode at the STM stage (accuracy within 5 K).

The surface was imaged with STM before Si deposition, then Si atoms (99.99% pure Si) of 0.01 ML were evaporated from a Knudsen cell [22] onto the sample at the sample temperature of ~125 K. The same areas were scanned to monitor surface morphology changes before and after the Si deposition. When the sample temperature was changed, we usually waited 20-30 min for the thermal stabilization of the sample so that the thermal drift did not affect STM imaging. In order to minimize the tip perturbation on the surface structure, the tunneling current was kept ≤ 0.2 nA.

### III. RESULTS AND DISCUSSION

#### A. Adsorption of single Si atoms on the ($\sqrt{7} \times \sqrt{3}$)-Pb phase.

Fig. 1(a-d) shows STM images of the ($\sqrt{7} \times \sqrt{3}$)-Pb phase acquired at four different sample biases when the sample temperature was cooled to 125 K. The images acquired at high biases exhibited one bright spot per unit cell. Typically, the empty-state image [Fig.

1(a)] exhibited a considerably higher image contrast than the filled-state image [Fig. 1(b)], even though the pattern appeared very similar for both polarities. For the images acquired at very low-biases, finer spots can be observed with each spot located roughly at the $T_1$ site [Fig. 1(c-d)], which is directly above a Si atom at the outmost layer of the Si(111) substrate. Each bright spot seen at high biases now appears as three smaller spots, a trimer, and the Pb trimers form into trimer rows along the $[\bar{1}\bar{1}2]$ direction. There are two additional spots per unit cell located between adjacent trimer rows. Notice that the image patterns are very similar for both polarities at low tunneling biases. Thus, it is very likely that the fine spots seen at low biases are related to the real atomic structure of Pb atoms at the surface, rather than the pure electronic structure. We note that this kind of fine structure can be resolved at a bias below a certain voltage, but the magnitude of the voltage varied from tip to tip. For filled-state images, the voltage can be about -1 V. For empty-state images, typically a lower voltage is required to resolve the fine structure.

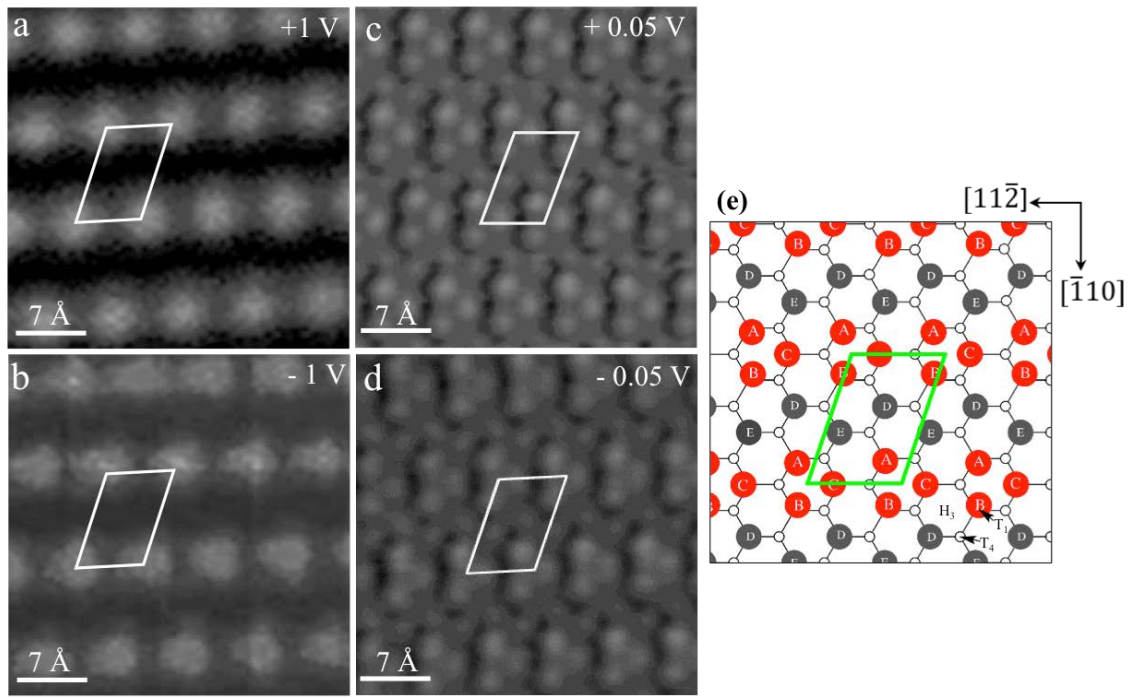

FIG. 1. (Color online) STM images of the Si(111)-($\sqrt{7}\times\sqrt{3}$)-Pb region acquired at different tunneling biases (a-d) and the atomic model (e). The sample bias is indicated at the upper-right hand corner. The sample temperature was 125 K. A unit cell of ($\sqrt{7}\times\sqrt{3}$) is outlined in each panel. Each bright spot seen at high tunneling biases can be resolved to be comprised of three spots (a Pb trimer) at low tunneling biases. There are five $T_1$ sites for each unit cell of ($\sqrt{7}\times\sqrt{3}$), which are indicated as $T_{1A}$, $T_{1B}$, $T_{1C}$, $T_{1D}$, and $T_{1E}$ sites, respectively. The Pb trimer (shown in red) is comprised of three Pb atoms at nearly $T_{1A}$, $T_{1B}$, and $T_{1C}$ sites. The arrows indicate two crystallographic directions of the Si(111) substrate. Surface high symmetric points ($T_1$, $T_4$ and $H_3$) are shown in the atomic model.

The atomic model of the ($\sqrt{7}\times\sqrt{3}$)-Pb phase based on the STM images acquired at low biases is illustrated in Fig. 1(e). There are five Pb atoms in a unit cell with each Pb atom located at nearly an on-top $T_1$ site. The Pb atoms are slightly displaced away from the exact $T_1$ site, but the illustrated displacements in Fig. 1(e) are exaggerated to make the appearance of each trimer more prominent. This model has a Pb coverage of 1 ML, corresponding well to previous coverage determination of monolayer Pb on Si(111) using RBS [14,18,24-28]. It has been shown that the reversible phase transition between the high-temperature (1×1)-Pb phase and the low-temperature ($\sqrt{7}\times\sqrt{3}$)-Pb phase does not involve any change in the Pb coverage [34,38]. In the literature, another model of the ($\sqrt{7}\times\sqrt{3}$)-Pb phase with the Pb coverage of 1.2 ML has been proposed based on X-ray diffraction [41] and *ab initio* calculations [42-46]. Here we adopted the trimer model [34,38] because the interpretation of our STM observations is simple and straightforward. We do not want to rule out the other model [41], so will discuss later.

In the model illustrated in Fig. 1(e), the five $T_1$ sites for each unit cell of ($\sqrt{7}\times\sqrt{3}$) are indicated as $T_{1A}$, $T_{1B}$, $T_{1C}$, $T_{1D}$, and $T_{1E}$ sites, respectively. The three Pb atoms at the neighboring $T_{1A}$, $T_{1B}$, $T_{1C}$ sites are centered at a three-fold hollow site ($H_3$ site), and can be viewed as a trimer. These three Pb atoms are shown in red to distinguish them from the Pb atoms at the other two $T_1$ sites. The Pb trimers form a row along the [$\bar{1}\bar{1}2$] direction. Since the Si(111) substrate has threefold symmetry, there are three possible row orientations.

Figures 2(a) and 2(b) show empty-state and filled-state images of the surface acquired before Si deposition at the sample temperature of 125 K, respectively. An area of the ($\sqrt{7}\times\sqrt{3}$)-Pb phase surrounded by a Si(111)-(7×7) reconstruction (the darker and more corrugated areas) was observed. After deposition of ~0.01 ML of Si, Si-induced point defects appeared as individual bright spots in the empty-state image [Fig. 2(c)] but dark defects in the filled-state images [Fig. 2(d)]. Higher-resolution images of the regions outlined in a white box in Figs. 2(c) and 2(d) were acquired again at lower biases, +1.0 V [Fig. 2(e)] and -1.0 V [Fig. 2(f)], respectively. Each Si-induced defect also appeared as a bright spot in the former case but as a missing defect near a $T_1$ site in a trimer structure in the latter. We note that there was no noticeable difference for empty-state images acquired between +1.0 and +1.5 V. Figures 2(g) and 2(h) are enlarged images within the white box in Figs. 2(e) and 2(f), respectively. For the same region, we acquired images at very low biases of +0.05 V [Fig. 2(i)] and -0.05 V [Fig. 2(j)]; the trimer structures are more evident. There are four Si-induced point defects on the surface as shown in Figs. 2(g) - 2(j). Three of them (numbers 1, 3, and 4) appeared as a dark defect near the $T_{1A}$ site, and one (number 2) near the $T_{1B}$ site. We acquired many STM images and observed dark defects associated with these two $T_1$ sites only and never with the $T_{1C}$, $T_{1D}$, or $T_{1E}$ sites. This indicates that the deposited Si atoms have a strong preference to occupy either the $T_{1A}$ or $T_{1B}$ site. Note that the empty-state image shown in Fig. 2(g) also reveals that the Si-induced bright spots are slightly shifted away from the center line of the trimer row. For defects 1, 3, and 4, the bright spots were shifted slightly upwards, toward the direction of the $T_{1A}$ site; for defect 2, the bright spot was shifted slightly downwards, toward the direction of the $T_{1B}$ site. Thus the empty-state images also allow us to determine whether the Si-induced effect is related to the $T_{1A}$ or $T_{1B}$ site. To achieve stable imaging of the surface structure and surface dynamics for an extended period of time and to obtain a high image contrast, we usually took empty-state images at high biases. It was actually very risky to image the ($\sqrt{7} \times \sqrt{3}$)-Pb surface at very low tunneling biases because the tip was very close to the sample and could have occasionally hit the surface.

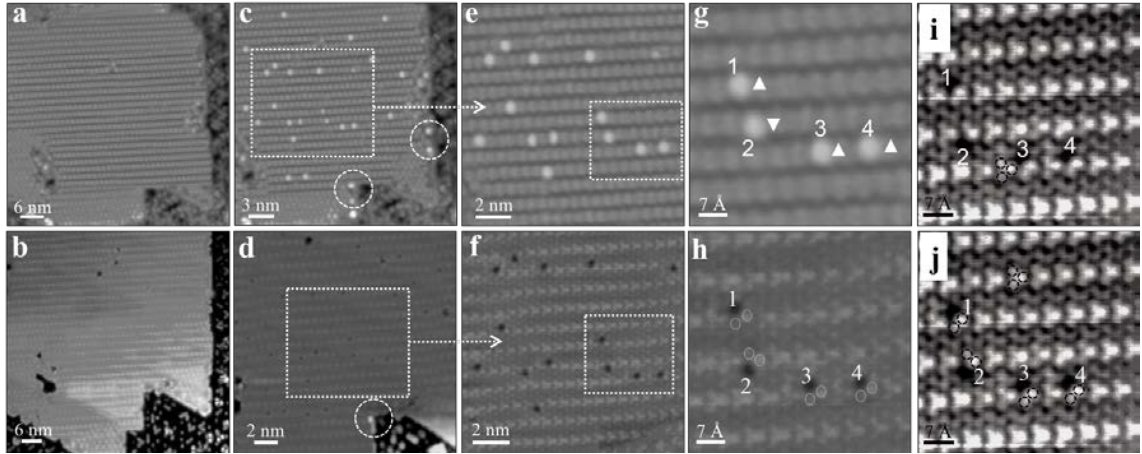

FIG. 2. STM images of a Pb-covered region before and after deposition of 0.01 ML of Si acquired at 125 K.  (a) Empty state (+1.5 V) and (b) filled state (-1.5 V) images before Si-deposition.  (c) Empty state (+1.5 V) and (d) filled state (-1.5 V) images of roughly the same area after Si deposition. Si deposition induced bright and dark spots in the empty- and filled-state images, respectively. Circles outline larger bright species, which are probably Si clusters, at the phase boundaries. (e) and (f) are the STM images of the same area inside a white box in (c) and (d) acquired at +1 V and -1 V, respectively. (g) and (h) are the enlarged images within the white box in (e) and (f) for better visualization of the detailed structure. Four Si-induced defects are numbered. The relative shift of the bright spots away from the center line of the trimer row is indicated with arrowheads in (g). In (h), the white circles outline the two spots of the original trimers for the Si-induced effects, which helps us identify whether the dark spots are related to $T_{1A}$ or $T_{1B}$ site. (i) and (j) are images acquired at +0.05 V and -0.05 V, respectively for the same region shown in (g) and (h).

## B. Hopping of individual Si atoms on the ($\sqrt{7} \times \sqrt{3}$)-Pb phase

The Si atoms were not mobile at 125 K, but they started to switch between the $T_{1A}$ and $T_{1B}$ sites within the same Pb trimer when the sample temperature was increased to ~152 K. Figure 3(a) show an image acquired at 150 K; the appearance and positions of the Si-induced bright spots did not change for over 30 min. We then increased the sample temperature to 152 K. The first image showed the slight shift of spot 3 [Fig. 3(b)], indicating its switching from the $T_{1A}$ to $T_{1B}$ site. For the image acquired 9 min and 24 min later, two more switches were identified for spots 1 and 2 [Figs. 3(c) and 3(d)].

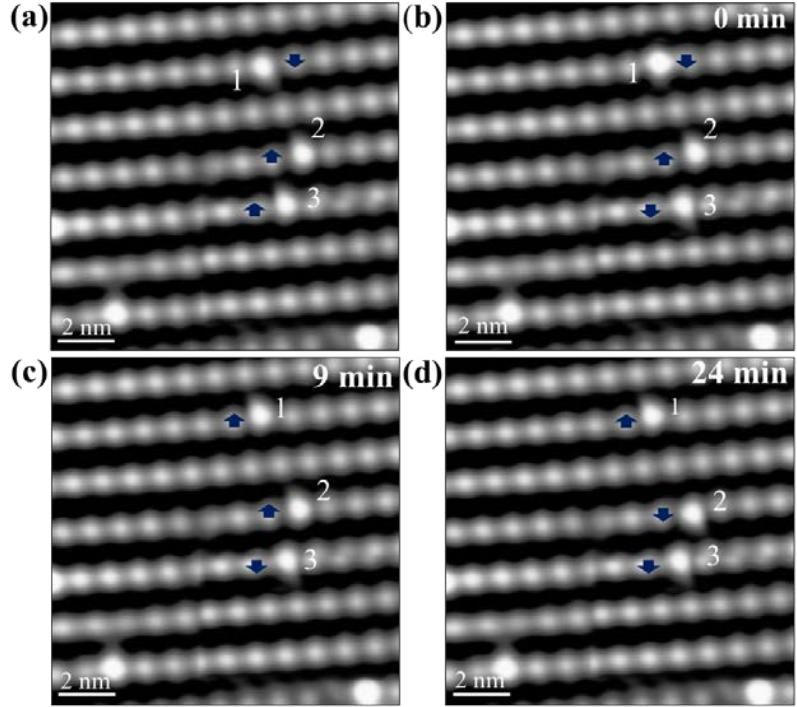

FIG. 3. (Color online) Switching of Si atoms between the $T_{1A}$ and $T_{1B}$ sites. (a) STM images of the adsorbed Si atoms on the ($\sqrt{7} \times \sqrt{3}$)-Pb surface at 150 K. (b)-(d) Switching of Si

atoms between the $T_{1A}$ and $T_{1B}$ sites after the sample temperature was raised to 152 K. The arrowheads denote the direction of the shift of the Si-induced bright spot: upward for the $T_{1A}$ site; and downward for the $T_{1B}$ site. The sample bias was +1.0 V. After the temperature was increased, we waited for 30-40 min to minimize the effect of thermal drift on the imaging. The time to acquire the first image in (b) was set to 0 min.

When the temperature was increased to ~160 K or higher, the Si atoms could hop away from their original trimer. We studied the hopping behavior of Si atoms from 160-174 K. The hops mainly occurred along the same trimer row; only two hops were observed away from a trimer row among more than 1000 hops. Figure 4 shows the hopping behaviors of Si atoms at 164 K. The arrowheads indicate the atom hopping direction and length of each hop. The Si atoms mainly hopped to a neighboring trimer (short hops) along the same trimer row, but occasionally hops with a length longer than one unit cell were observed. Fig. 5 shows the hopping of Si atoms at 170 K. Evidently, the hopping rate of Si atoms increased with sample temperature, indicating that the hopping is a thermally activated process. An Arrhenius plot for the hopping of Si atoms along the trimer row is shown in Fig. 6. An activation energy of ~0.41 eV and a prefactor of ~$10^{11}$ s$^{-1}$ were derived.

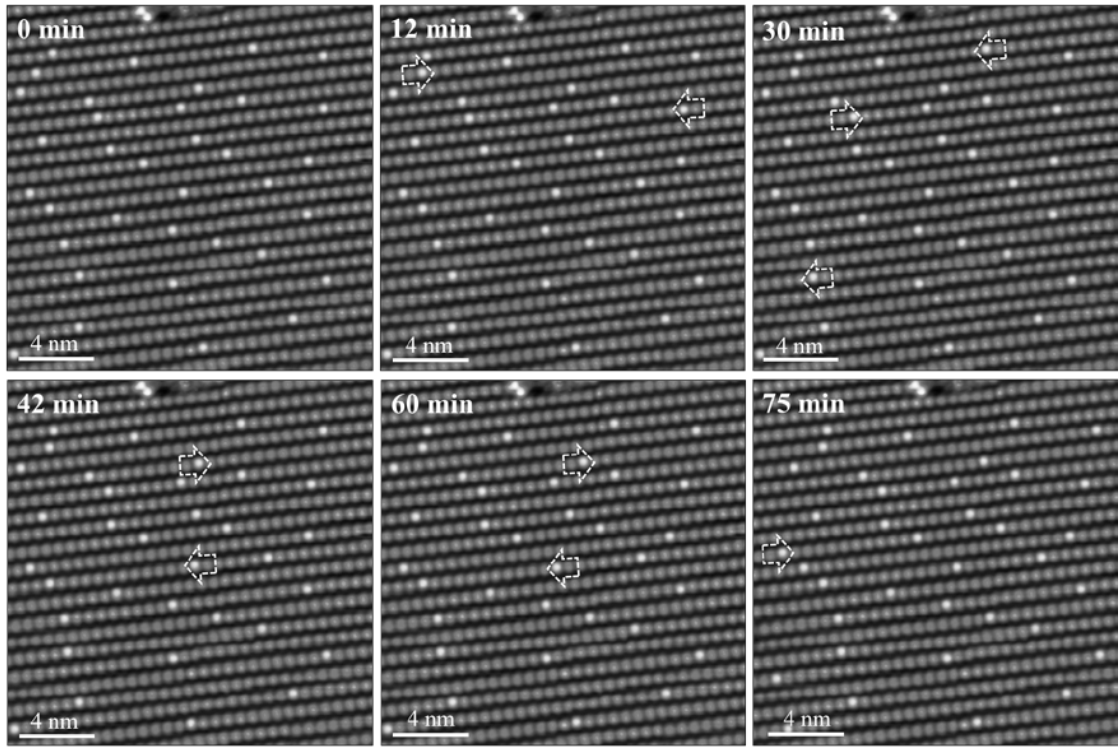

FIG. 4. STM images showing the site hopping of Si atoms along the trimer row at 164 K acquired at the sample bias of +1.0 V. It took 3 min to acquire one image; the time is indicated at the upper-left hand corner of each panel. The white dotted arrowheads indicate the direction and length of each hop. Mainly short hops (i.e. hops to a neighboring trimer along the same row) were observed at this temperature.

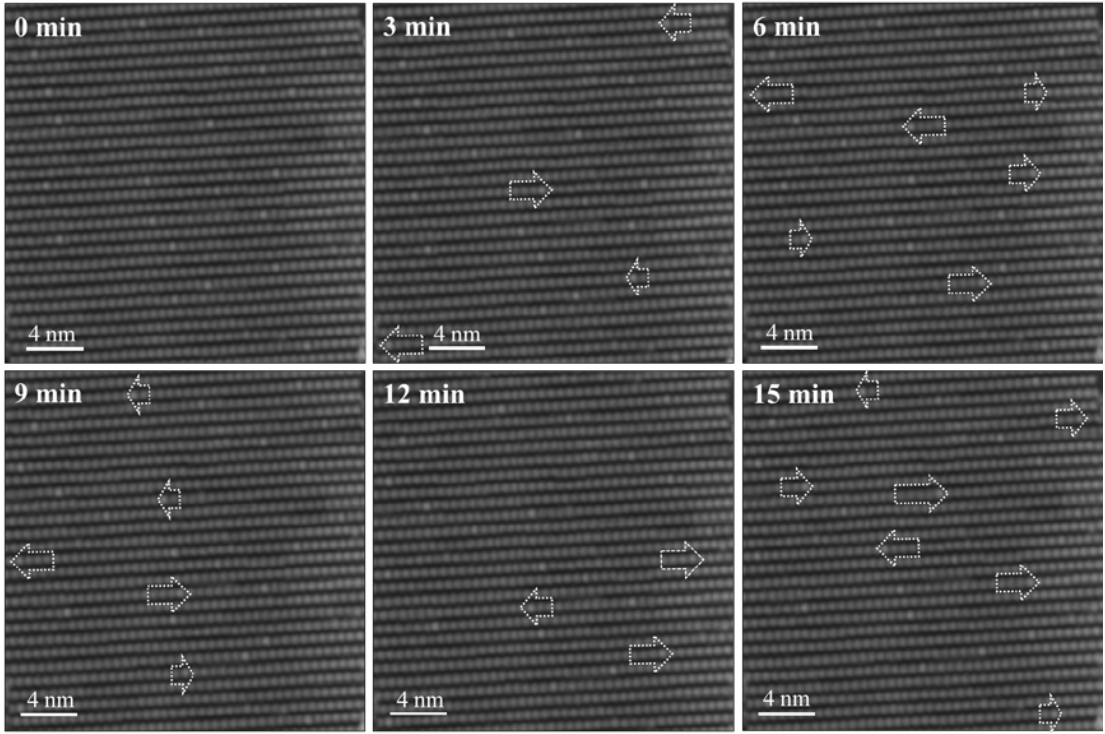

FIG. 5. STM images showing the site hopping of Si atoms along the trimer row at 170 K acquired at the sample bias of +1.5 V. It took 3 min to acquire one image; the time is indicated in each panel. The white dotted arrows indicate the direction and length of each hop. Both long and short hops were observed.

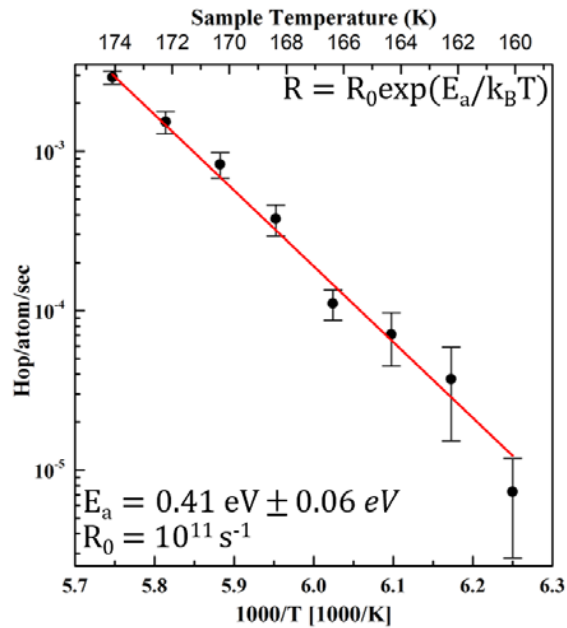

FIG. 6. (Color online) Arrhenius plot of the hopping of single Si atoms on the ($\sqrt{7} \times \sqrt{3}$)-

Pb surface from 160-174 K. The activation energy $E_a$ and the pre-exponential factor $R_0$ were obtained from the best fit line (solid red line) to the Arrhenius expression. The calculation of the hopping rate included both the short and long hops.

Another interesting observation is that the ratio of long hops to short hops increases dramatically with sample temperature. Figure 7(a) shows the hopping rates of short, double, long, and total hops (in terms of hops per atom per frame) at different temperatures. Here we separate long hops (i.e. hops to a distance of three unit cells or longer) from double hops (hops to a distance of two unit cells) because a portion of double hops might be caused by two independent short hops within 3 min of our image acquisition time. Figure 7(b) shows the percentage of short, double, and long hops as a function of temperature. The probability of long hops being caused by independent short hops is negligible. Short hops dominated at low temperatures. The rates of short, double, and long hops basically increased with temperature; but evidently, the rate of long hops increased much more rapidly with temperature than the other two. Long hops dominated at the temperatures above 170 K.

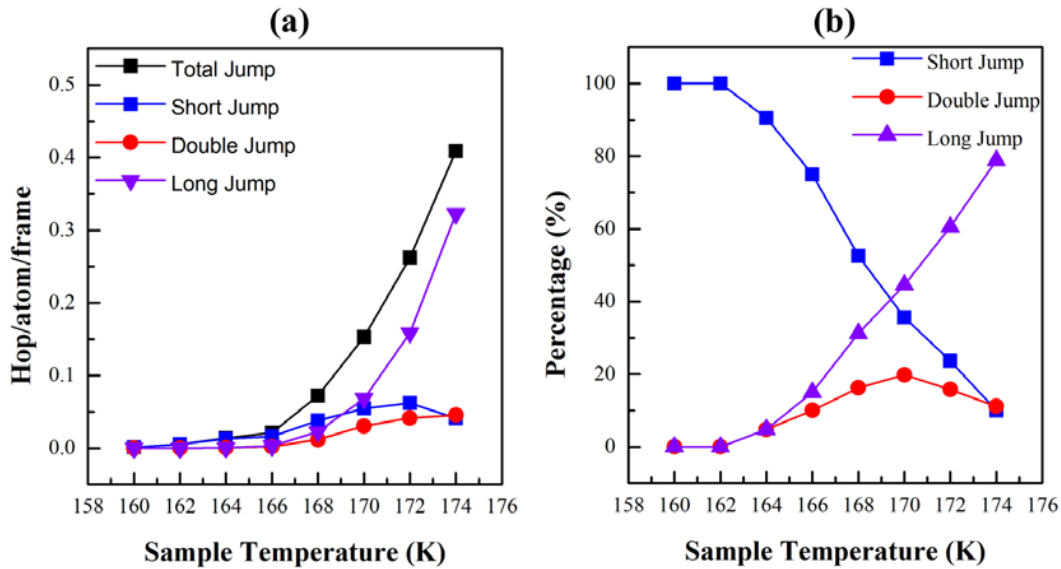

FIG. 7. (Color online) (a) Rate of short, double, long and total hops of single Si atoms on the (√7x√3)-Pb surface at different temperatures. (b) Percentage of short, double, and long hops at different temperatures.

### C. Breaking of mirror symmetry in the structure of the ($\sqrt{7}\times\sqrt{3}$)-Pb phase

Another interesting observation is that Si atoms have an elevated tendency to occupy the $T_{1A}$ site relative to the $T_{1B}$ site. Figures 8(a) and 8(b) show filled-state images acquired at 125 K and 160 K, respectively. The Si-induced dark sites mainly appeared near to the $T_{1A}$ site. We analyzed many Si-induced defects from 125-176 K and calculated the ratio of defects that appears near the $T_{1A}$ site to those near the $T_{1B}$ site [Fig. 8(c)]. At temperatures, below 150 K, when the dark sites did not switch between the $T_{1A}$ and $T_{1B}$ sites, the ratio of dark species near the $T_{1A}$ site relative to the $T_{1B}$ site was 4.6-5.0. At temperatures above 150 K, when switching between the two sites became possible, the ratio reduced slightly to 4.0-4.8. The ratios were considerably higher than 1, indicating that the Si had a lower adsorption potential near the $T_{1A}$ site than the $T_{1B}$ site. We estimated the adsorption potential difference as 20 ± 2 meV. Previous atomic model [34,38,41,42] proposed for the ($\sqrt{7}\times\sqrt{3}$)-Pb phase possess mirror symmetry, i.e. the $T_{1A}$ and $T_{1B}$ sites are energetically equivalent. This study provides evidence for the breaking of the mirror symmetry, which has major implications on the atomic structure of this phase. Further discussion of the atomic structure of the ($\sqrt{7}\times\sqrt{3}$)-Pb phase can be found in Section III-F.

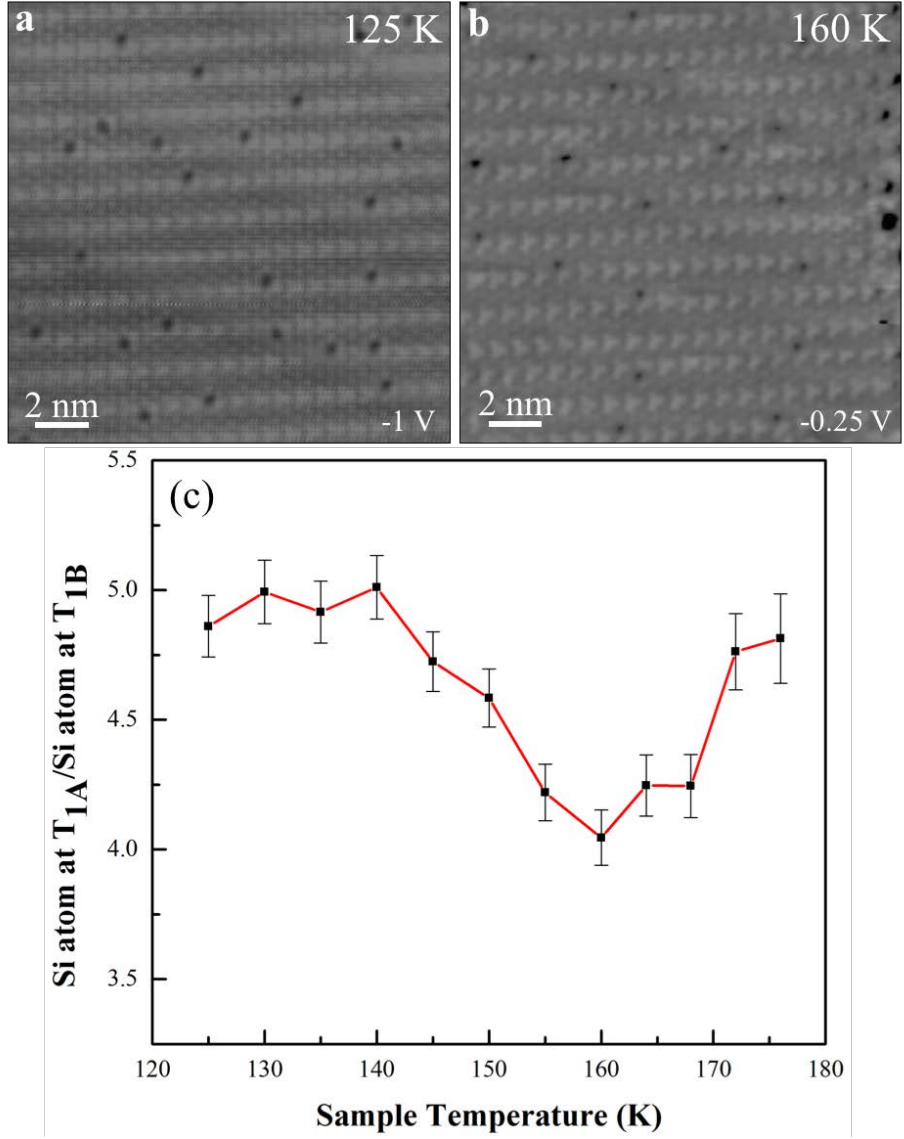

FIG. 8. (Color online) Anisotropic adsorption behavior of Si atoms on the ($\sqrt{7} \times \sqrt{3}$)-Pb surface. (a) Filled-state image acquired at 125 K and sample bias of -1.0 V. (b) Filled-state image acquired at 160 K and sample bias of -0.25 V. (c) Ratio of Si-induced defects near the $T_{1A}$ site to those near the $T_{1B}$ site.

### D. Aggregation of Si atoms into Si clusters at temperatures above ~180 K

During our observation of the hopping of Si atoms on the ($\sqrt{7} \times \sqrt{3}$)-Pb surface, we did not observe any aggregation of neighboring Si atoms into a cluster, suggesting repulsive interactions between single Si atoms. However, at temperatures above ~170 K, when long hops of Si atoms dominated, we observed the appearance of Si clusters at phase boundaries as well as the decrease in the number density of single Si atoms. In another set of experiments, we monitored the same area of the surface at different temperatures after Si deposition at temperatures ~125 K. An example is shown in Fig. 9. Figure 9(a) shows an empty-state image acquired at 168 K: several single Si atoms (the bright species) were evident on the surface. After increasing the sample temperature to 172 K [Fig. 9(b)], the number of single Si atoms significantly decreased; meanwhile, we observed the appearance of a few large bright species, presumably Si clusters, at phase boundaries or defective sites. The number of Si atoms decreased further after increasing the temperature to 175 K [Fig. 9(c)], meanwhile, more clusters were seen. At 181 K, bright species corresponding to the adsorption of single Si atoms disappeared completely, and some Si clusters seemed to change sites [Fig. 9(d)]. The disappearance of single Si atoms might be attributed to a very high hopping rate at such a high temperature, but this possibility can be ruled out because they did not reappear after the sample was cooled down to 155 K, when the hopping of Si atoms did not occur [Fig. 9(e)].

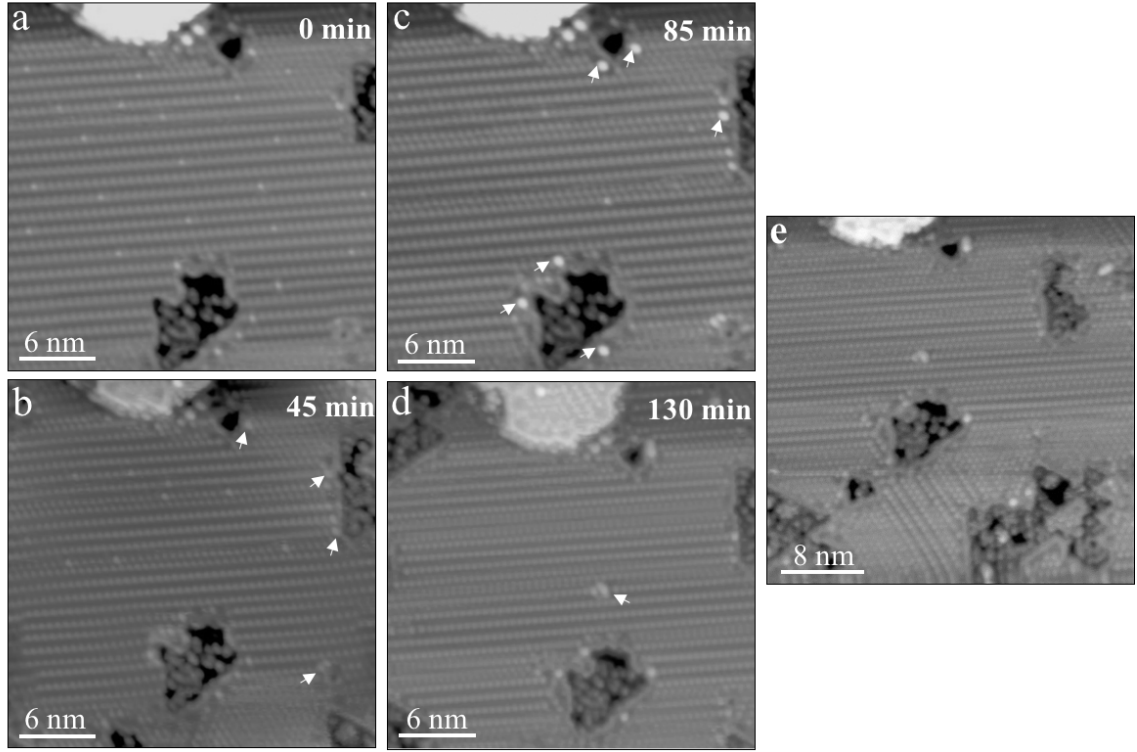

FIG. 9. Empty-state images of roughly the same area on the ($\sqrt{7} \times \sqrt{3}$)-Pb surface acquired at the sample temperature of (a) 168 K, (b) 172 K, (c) 175 K, (d) 181 K, and finally at (e)155 K. The sample bias was +1.25 V. The number density of single Si atoms decreased with increasing temperature; meanwhile, the appearance of Si clusters (some are indicated with white arrowheads) was seen at phase boundaries or defective sites. The darker areas are the remnant regions of Si(111)-(7x7) that were not converted into the monolayer Pb covered regions during the sample preparation. The bright area at the top of each image is a Pb-covered Si(111) island of one bilayer thickness, which often appears during the annealing process [22,34,38,40].

We also deposited Si at 230 K and then cooled the samples to 140-170 K; only large bright species (Si clusters) were observed at phase boundaries or defective sites, and no species related to adsorption of single Si atoms was observed (data not shown). We thus conclude that the deposited Si atoms had all aggregated into Si clusters at temperatures above ~180 K, and the transition was irreversible. As epitaxial growth of Si thin films is typically carried out at temperatures above 300 K, one can expect that Si clusters, rather than individual Si atoms, play the major role in the nucleation and growth of epitaxial thin films. This has important implications on the atomic mechanism of epitaxial growth, because conventional nucleation theories do not consider such a scenario. Recently, a new scenario based on the presence and reactions of magic clusters was proposed for epitaxial growth of covalent materials [40]. The current study provides experimental evidence to support such a scenario.

### E. Hopping mechanism of single Si atoms on the ($\sqrt{7} \times \sqrt{3}$)-Pb surface

Our experiments indicated that single Si atoms appeared only near the $T_{1A}$ and $T_{1B}$ sites but not near the other three $T_1$ sites in the temperature range of 125-176 K. Evidently, the $T_{1A}$ and $T_{1B}$ sites are the low-energy sites for the adsorption of single Si atoms. The switching of single Si atoms between the $T_{1A}$ and $T_{1B}$ sites in the same trimer occurred at lower temperatures than the hopping between different trimers, suggesting that switching has a lower activation energy than hopping. The observations that single Si atoms mainly hopped between trimers on the same trimer row but rarely hopped across the trimer row are very interesting. We postulate that the Si atoms probably hop through the $T_{1C}$ site in order to hop to the $T_{1A}$ or $T_{1B}$ site of a neighboring trimer. The $T_{1C}$ site may have a higher adsorption potential than the $T_{1A}$ and $T_{1B}$ sites. The activation energy for hopping from the $T_{1C}$ site to a neighboring $T_{1A}$ or $T_{1B}$ site is probably low so that the lifetime for the adsorption of Si atom near the $T_{1C}$ site is shorter than the data acquisition time (~1 ms) in our STM imaging. The activation energy derived in Fig. 6 may be related to the energy barrier for single Si atoms to hop from a $T_{1A}$ or $T_{1B}$ site to a neighboring $T_{1C}$ site.

The observations that the hopping of single Si atoms mainly occurred along the same trimer row indicate that there is a high energy barrier for single Si atoms to hop across to a neighboring trimer row. The adsorption potential near the $T_{1D}$ or $T_{1E}$ site is likely even higher than the $T_{1C}$ site; thus, the energy barrier for single Si atoms to hop from the $T_{1A}$ or $T_{1B}$ site to a neighboring $T_{1D}$ or $T_{1E}$ site is too high in the temperature range of our experiments.

It is interesting to observe the sharp increase in the percentage of long hops at temperatures higher than ~170 K. A new hopping pathway, which does not need to go through the $T_{1A}$ or $T_{1B}$ site of the neighboring trimer, could become possible at higher sample temperature. Long hops of atoms or clusters have been reported in several previous studies using field ion microscope or STM [47-51]; some also show that short hops dominate at low temperatures but long hops dominate at high temperatures [51]. The behavior of long hops is an interesting phenomenon in surface diffusion. Our observations of formation of Si clusters occuring at temperatures that long hops dominated suggest that that Si atoms hopped to the phase boundaries or defective sites and became trapped there, and the aggregation of Si atoms into clusters then occurred.

## F. Atomic structure of the Si(111)-($\sqrt{7} \times \sqrt{3}$)-Pb phase

There has been some controversy about the atomic structure of the ($\sqrt{7} \times \sqrt{3}$)-Pb phase. Here we adopt the trimer model [34,38] because it is more straightforward to interpret our experimental data. The trimer model was proposed based on STM images acquired at low biases. In this model, all Pb atoms are located near the $T_1$ site, and the three neighboring Pb atoms around the $T_{1A}$, $T_{1B}$, and $T_{1C}$ sites form a trimer. Another model has been proposed based on X-ray diffraction [41] and *ab-initio* calculations [42-46] [Fig. 10(a)]. The Pb coverage is 6/5 ML, which has an extra Pb atom for each ($\sqrt{7} \times \sqrt{3}$) unit cell compared to the trimer model [34,38]. This extra Pb atom is located at the center of the Pb trimer, an $H_3$ site [atom 1 in Fig. 10(a)]. Due to the presence of this extra Pb atom, the three surrounding Pb atoms [i.e. atoms 2, 3, and 6 in Fig. 10(a)] are displaced somewhat outward from their related $T_1$ sites towards adjacent $T_4$ sites. This is also reflected in the calculated STM images [42]. There are two Pb atoms (atoms 3 and 4) corresponding to the Pb atoms in the $T_{1D}$ and $T_{1E}$ sites in the trimer model [Fig. 1(e)]; they are also displaced slightly from the $T_{1D}$ and $T_{1E}$ sites.

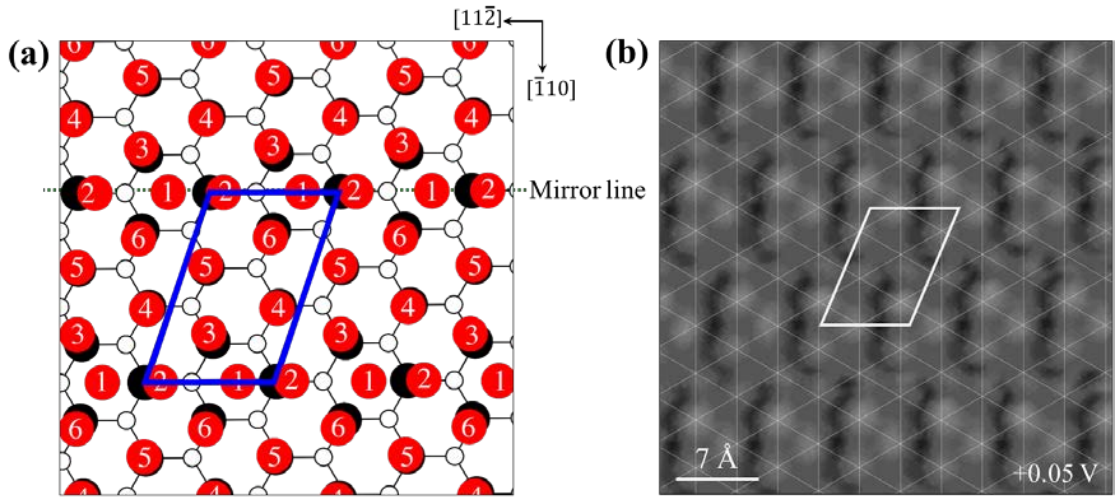

FIG. 10. (Color online) (a) Atomic model of the ($\sqrt{7} \times \sqrt{3}$)-Pb phase with the Pb coverage of 6/5 ML proposed based on an X-ray diffraction study [41]. (b) STM image of the (√7x√3)-Pb phase overlain with a grid with interceptions at the $T_1$ sites. The white lines are along the Si(111) substrate orientations, $\{\bar{1}\bar{1}2\}$.

Fig. 10(b) shows a STM image of the ($\sqrt{7} \times \sqrt{3}$)-Pb phase overlaid with lines connecting the $T_1$ sites in three directions. The lines form a lattice grid with interceptions at the $T_1$ sites [Fig. 10(b)]. One can see that all the bright spots in STM images are very close to the exact $T_1$ sites. We do not see the extra spot at the center of the Pb trimer, an $H_3$ site, which in fact exhibited a depression in our images acquired at low biases. In addition, the bright spots corresponding to the Pb trimer do not exhibit outward displacement away from the $T_1$ site. The bright spot associated with the $T_{1C}$ site even exhibited a small displacement toward the center $H_3$ site, contrary to the outward displacement in the calculated STM image for the model of 6/5 ML Pb. We checked STM images of the ($\sqrt{7} \times \sqrt{3}$)-Pb phase acquired at low biases from different experiments using several different STM tips. All exhibited a similar feature to that shown in Fig. 10(b).

Since we observed a small difference in the adsorption potential for single Si atoms on the $T_{1A}$ and $T_{1B}$ sites, we believe the atomic model for the ($\sqrt{7} \times \sqrt{3}$)-Pb phase should show breaking in the mirror symmetry of Pb atoms near $T_{1A}$ and $T_{1B}$ sites. Thus, the Pb atoms near the $T_{1A}$ and $T_{1B}$ sites should have different displacements (even though they may be very small) from their related $T_1$ site. The model proposed based on X-ray diffraction [41] and *ab-initio* calculations [42] possesses a mirror symmetry on both sides of the mirror line [Fig. 10(a)]. This model should also be modified to break the mirror symmetry. Therefore, further experimental and theoretical calculations will be needed to determine the atomic structure precisely.

## IV. CONCLUSIONS

We have studied the very initial stage of Si adsorption on the Si(111)-($\sqrt{7} \times \sqrt{3}$)-Pb surface at temperatures from 125-230 K. We observed that the deposited Si atoms were adsorbed near two specific $T_1$ sites ($T_{1A}$ and $T_{1B}$) only among the five $T_1$ sites in a unit cell of the ($\sqrt{7} \times \sqrt{3}$)-Pb surface at low temperature (~125 K). The adsorption also exhibited strong anisotropy between the $T_{1A}$ and $T_{1B}$ sites, indicating that these two sites are not energetically equivalent. We thus conclude that the ($\sqrt{7} \times \sqrt{3}$)-Pb structure does not possess the mirror symmetry as assumed in previously proposed atomic models. When Si atoms were deposited on this surface, single Si atoms were stationary at the temperatures below ~150 K, but they were found to switch between the $T_{1A}$ and $T_{1B}$ sites when the sample temperature exceeded ~150 K. When the sample temperature was increased above ~160 K, the Si atoms hopped along the trimer row. Short hops dominated at the temperature below 170 K but long hops dominated at higher temperatures. An energy barrier of ~0.41 eV and a preexponential factor of $10^{11}$ s were derived based on continuous-time STM imaging of the hopping events acquired from 160-174 K. At the temperature above ~170 K, Si atoms may hop to phase boundaries or defective sites and aggregate into Si clusters. At a temperature of ~180 K, no single Si atoms were observed and only Si clusters were present. As the epitaxial growth of semiconductor systems typically occurs at temperatures above room temperature, we expect that clusters, rather than single deposited atoms (or monomers), play the major role in nucleation and growth mechanisms. This may have major implications on the atomic mechanism of epitaxial growth of covalent materials because current nucleation theories do not consider such a scenario. A new scenario may be required to understand the mechanism of epitaxial growth of covalent materials.

**ACKNOWLEDGEMENTS**

This work was supported by Academia Sinica.

**REFERENCES AND FOOTNOTES**

\* Author to whom correspondence should be addressed.

Email  ishwang@phys.sinica.edu.tw